\documentclass[longauth]{aa} 
\usepackage{graphicx}
\usepackage{txfonts}
\usepackage{natbib}
\usepackage{multirow}
\newcommand{\mia}{Sh2-104}
\newcommand{\hii}{H\,{\sc ii}}
\newcommand{\nii}{$[$N{\sc ii}$]\;^3P_1-^3P_0$}
\newcommand{\ci}{$[$C{\sc i}$]\;^3P_1-^3P_0$}
\newcommand{\cia}{$[$C{\sc i}$]\;^3P_2-^3P_1$}
\begin{document}

\title{Physical properties of the \mia\ \hii\ region as seen by
\textit{Herschel}\thanks{\textit{Herschel} is an ESA space observatory with
science instruments
provided by European-led Principal Investigator consortia and with important
participation from NASA.}}

\author{
  J. A. Rod\'on\inst{1}\and
  A. Zavagno\inst{1}\and
  J.-P. Baluteau\inst{1}\and
  L. D. Anderson\inst{1}\and
  E. Polehampton\inst{2,3}\and
  A. Abergel\inst{4}\and
  F. Motte\inst{5}\and
  S. Bontemps\inst{5,6}\and
  P. Ade\inst{17}\and
  P. Andr\'e\inst{5}\and
  H. Arab\inst{4}\and
  C. Beichman\inst{8}\and
  J.-P. Bernard\inst{7}\and
  K. Blagrave\inst{14}\and
  F. Boulanger\inst{4}\and
  M. Cohen\inst{9}\and
  M. Compiegne\inst{14}\and
  P. Cox\inst{10}\and
  E. Dartois\inst{4}\and
  G. Davis\inst{11}\and
  R. Emery\inst{17}\and
  T. Fulton\inst{20}\and
  C. Gry\inst{1}\and
  E. Habart\inst{4}\and
  M. Halpern\inst{13}\and
  M. Huang\inst{11}\and
  C. Joblin\inst{7}\and
  S. C. Jones\inst{2}\and
  J. Kirk\inst{17}\and
  G. Lagache\inst{4}\and
  T. Lin\inst{3}\and
  S. Madden\inst{5}\and
  G. Makiwa\inst{2}\and
  P. Martin\inst{14}\and
  M.-A. Miville-Desch\^enes\inst{4}\and
  S. Molinari\inst{15}\and
  H. Moseley\inst{19}\and
  D. Naylor\inst{2}\and
  K. Okumura\inst{5}\and
  F. Orieux\inst{12}\and
  D. Pinheiro Gon\c{c}alvez\inst{14}\and
  T. Rodet\inst{12}\and
  D. Russeil\inst{1}\and
  P. Saraceno\inst{15}\and
  S. Sidher\inst{3}\and
  L. Spencer\inst{2}\and
  B. Swinyard\inst{3}\and
  D. Ward-Thompson\inst{17} \and
  G. White\inst{18,21}
  }

  \institute{
  Laboratoire d'Astrophysique de Marseille (UMR 6110 CNRS \&
  Universit\'e de Provence), 38 rue F. Joliot-Curie, 13388 Marseille
  Cedex 13, France.  \email{jarodon@oamp.fr} \and
 Institute for Space Imaging Science, University of Lethbridge, Lethbridge,
Canada \and
 Space Science Department, Rutherford Appleton Laboratory, Chilton, UK \and
  Institut d'Astrophysique Spatiale, CNRS/Universit\'e Paris-Sud 11, 91405
  Orsay, France  \and
   CEA, Laboratoire AIM, Irfu/SAp, Orme des Merisiers, F-91191
Gif-sur-Yvette, France \and
 CNRS/INSU, Laboratoire d'Astrophysique de Bordeaux, UMR 5804, BP 89, 33271
Floirac cedex, France \and
 Universi\'e de Toulouse ; UPS ; CESR ; and CNRS ; UMR5187,
 9 avenue du colonel Roche, F-31028 Toulouse cedex 4, France \and
 Infrared Processing \& Analysis Center, California Institute of
Technology, Mail Code 100-22, 770 South Wilson Av, Pasadena, CA 91125,  USA \and
 University of California, Radio Astronomy Laboratory, Berkeley, 601 Campbell
Hall, US Berkeley CA 94720-3411, USA \and
Institut de Radioastronomie Millim\'etrique (IRAM), 300 rue de la Piscine,
F-38406 Saint Martin d'H\`eres, France  \and
National Astronomical Observatories (China) \and
Laboratoire des Signaux et Syst\`emes (CNRS \& Sup\'elec \& Universit\'e
Paris-Sud 11), Moulon, 91192 Gif-sur-Yvette, France \and
 Department of Physics and Astronomy, University of British Columbia, Vancouver,
Canada \and
 Canadian Institute for Theoretical Astrophysics, Toronto, Ontario, M5S 3H8,
Canada  \and
 Istituto di Fisica dello Spazio Interplanetario, INAF, Via del Fosso
del Cavaliere 100, I-00133 Roma, Italy \and
 Dept. of Physics \& Astronomy, University College London,
Gower Street, London WC1E 6BT, UK \and
Department of Physics and Astronomy, Cardiff University, UK \and
 Centre for Astrophysics and Planetary Science, School of Physical Sciences,
University of Kent, Kent, UK \and
 NASA - Goddard SFC, USA\and
Blue Sky Spectrosocpy Inc, Lethbridge, Canada\and
Department of Physics \& Astronomy, The Open University, Milton Keynes MK7 6AA,
UK
             }

   \date{Received March 31, 2010 ; accepted May 7, 2010}

 
  \abstract
   {\mia\ is a Galactic \hii\ region with a bubble morphology, detected at
optical and radio wavelengths. It is considered the first observational
confirmation of the collect-and-collapse model of triggered star-formation. }
   {We aim to analyze the dust and gas properties of the \mia\ region to
better constrain its effect on local future generations of stars. In
addition, we investigate the relationship between the dust emissivity index
$\beta$ and the dust temperature, $T_{dust}$.}
   {Using \textit{Herschel} PACS and SPIRE images at $100$, $160$, $250$,
$350$ and $500\,\mu$m
   we determine $T_{dust}$ and $\beta$ throughout \mia,
fitting the spectral energy distributions (SEDs) obtained from aperture
photometry. With the SPIRE Fourier transform spectrometer (FTS) we obtained
spectra at different positions in the \mia\ region.
We detect $J$-ladders of $^{12}$CO and $^{13}$CO, with which we derive the gas
temperature and column density. We also detect proxies of ionizing flux as the
\nii\ and \cia\ transitions.}
   {We find an average value of $\beta \sim 1.5$ throughout \mia, as well as a
$T_{dust}$ difference between the photodissociation region (PDR, $\sim25\,$K)
and the interior ($\sim40\,$K) of the bubble. We recover the anti-correlation
between $\beta$ and dust temperature reported numerous times in the literature.
The relative isotopologue abundances of CO appear to be enhanced above
the standard ISM values, but the obtained value is very preliminary and
is still affected by large uncertainties.
}
   {}


\authorrunning{J. A. Rod\'on et al.}

   \maketitle
%

\section{Introduction}

Sharpless 104 (\mia, \citealt{sharpless1959}) is an optically
visible Galactic \hii\ region with a bubble morphology, excited by an
O6V star \citep{crampton1978, lahulla1985}. It is located $\sim4\,$kpc from
the Sun \citep{deharveng2003}, with galactic coordinates $74.7620;
+0.60$ (J2000).

\citet{deharveng2003} proposed \mia\ as a strong candidate for massive
triggered star formation through the collect-and-collapse process
\citep{elmegreen1977}.
The ionized region is also visible at radio wavelengths \citep{fich1993},
and an
ultracompact (UC) \hii\ region, coincidant with the IRAS~20160+3636 source, lies
at its eastern border (\citealt{condon1998}).

We present new submm images and spectra taken towards \mia\ with
the \textit{Herschel} Space Observatory \citep{pilbratt2010}.
These observations allow us to map a wavelength range not easily accessed
before, providing new insights into the dust and gas properties of \mia.

\section{Observations}

The \textit{Herschel} observations were taken on 2009 December 17
simultaneously with
PACS \citep{poglitsch2010} and SPIRE \citep{griffin2010}, as
part of the guaranteed-time key-projects ``Evolution of Interstellar Dust'' of
SPIRE \citep{abergel2010}, and
HOBYS of PACS \citep{motte2010}. A $15'\times15'$ region was imaged with
PACS at $100$ and $160\,\mu$m (at resolutions of $10''$ and $14''$), and with
SPIRE at $250$, $350$ and $500\,\mu$m (at resolutions of $18''$, $25''$ and
$36''$). Spectra were taken with the SPIRE-FTS long and short wavelength
receivers (SLW and SSW, respectively) at seven different positions with sparse
sampling, covering the $194-671\,\mu$m range. The resolution at the receivers'
central pixels varies between $17-19''$ for SSW and $29-42''$ for SLW.
The data were reduced with the HIPE software version 2.0 with the
latest standard calibration \citep{swinyard2010}.

\begin{figure}[h]
  \centering
  \includegraphics[width=\columnwidth]{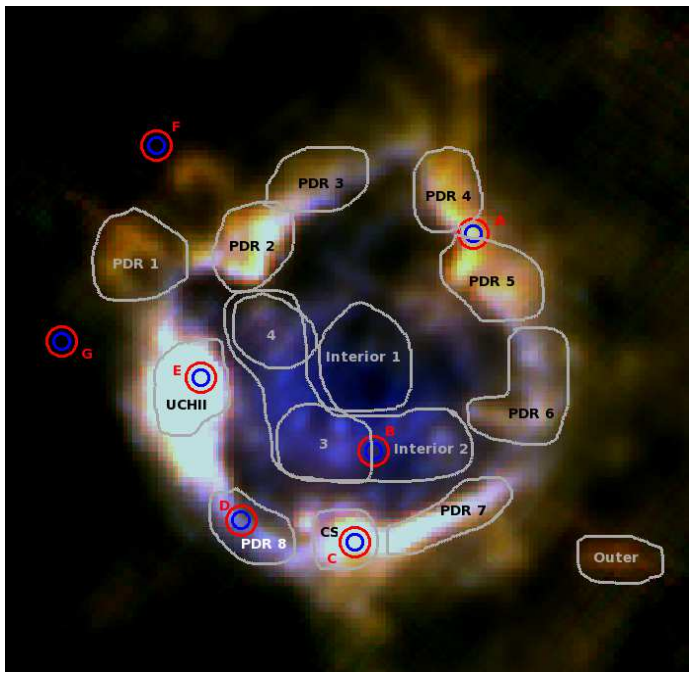}
  \caption{Composite image of \mia. The field is $\sim13'\times13'$ and
shows PACS $100\,\mu$m emission in blue, SPIRE
$250\,\mu$m in green and SPIRE $500\,\mu$m in red. Outlined are the
regions where aperture photometry was applied, the region used for
background substraction is off the map. Blue and red circles mark the
pointings of the central pixels of SPIRE-FTS SSW and SLW, respectively. Their
diameter represents their respective average resolution of $19''$ and $35''$.}
  \label{fig-sh104-main}
\end{figure}

Figure \ref{fig-sh104-main} shows a color-composite image of \mia\ with PACS
$100\,\mu$m (blue), SPIRE $250\,\mu$m (green) and SPIRE $500\,\mu$m (red).
Different regions of interest, addressed in the following sections, are
superimposed.
We can see that the interior of the bubble is brighter in the PACS band, showing
the hotter temperatures of the material in this region. On the other hand,
outside the bubble the material is colder and emits stronger in the SPIRE
bands.

\section{Dust properties}

We assume the dust emission in \mia\ can be modeled by
an (optically thin) gray-body and that the emissivity of the dust
grains can be fitted with a power law 
\begin{equation}
  S_{\nu} \propto \Omega B_{\nu} \left(T\right)\kappa_{0}
\left(\frac{\nu}{\nu_{0}}\right)^{\beta}N,
\end{equation}
where $S_{\nu}$ is the measured flux density at frequency $\nu$, 
$\Omega$ is the observing beam solid angle, 
$B_{\nu}\left(T\right)$ is the Planck function for temperature $T$ at frequency
$\nu$, $\kappa_{0}\left(\nu/\nu_0\right)^{\beta}$ is the
dust opacity, $\beta$ is the dust emissivity index, and $N$ is the dust column
density \citep{hildebrand1983}.
The value of $\beta$ is believed to range between $\sim1$ and $\sim2$, but is an
open issue that is still discussed, as is its dependence on the dust
temperature (see e.g., \citealt{ossenkopf1994}). \citet{dupac2003} and
\citet{desert2008} found an inverse relationship between $\beta$ and $T_{dust}$.
On the other hand, \citet{shetty2009} suggested that this result arises from
noise and the combination of multiple emission components along the line of
sight.

\begin{figure}[h]
  \centering
  \includegraphics[width=\columnwidth]{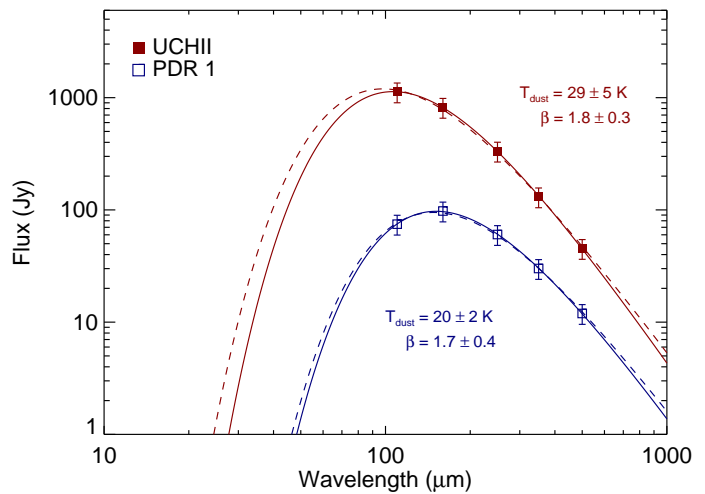}
  \caption{Example of the SEDs obtained for regions \textit{UCHII} (filled red
squares) and PDR 1 (open blue squares). A dashed line is the fit with $\beta =
1.5$ and a solid line is the fit with $\beta$ allowed to vary, with the
resulting $(T_{dust};\beta)$ values shown in the respective color. The
temperature
difference between the regions is apparent, as is the difference in the
peak sampling of the SED for \textit{PDR 1} and \textit{UCHII}.}
  \label{fig-seds}
\end{figure}

\begin{table}[b]
  \renewcommand{\arraystretch}{1.1}  
  \begin{minipage}{\columnwidth}
    \centering
    \caption{Results from aperture photometry.}
    \label{table-beta}
    \renewcommand{\footnoterule}{}
    \begin{tabular}{l|cc|c}
      \hline\hline
      \multirow{2}{*}{Region} & T$_{dust}$ (K) & $\beta$ & T$_{dust}$ (K) \\
       & \multicolumn{2}{c|}{$\beta$ Free} & $\beta=1.5$ \\
      \hline
      CS & 25 $\pm$ 3 & 1.5 $\pm$ 0.3 & 25.5 $\pm$ 0.9 \\
      Interior 1 & 43 $\pm$ 12 & 1.2 $\pm$ 0.4 & 35.7 $\pm$ 2.1 \\
      Interior 2 & 39 $\pm$ 9 & 1.3 $\pm$ 0.3 & 33.5 $\pm$ 1.7 \\
      Interior 3 & 47 $\pm$ 15 & 1.0 $\pm$ 0.4 & 34.2 $\pm$ 1.9 \\
      Interior 4 & 35 $\pm$ 8 & 1.3 $\pm$ 0.4 & 30.4 $\pm$ 1.4 \\
      Outer & 17 $\pm$ 2 & 1.7 $\pm$ 0.4 & 17.3 $\pm$ 0.4 \\
      PDR 1 & 20 $\pm$ 2 & 1.7 $\pm$ 0.4 & 21.7 $\pm$ 0.6 \\
      PDR 2 & 22 $\pm$ 3 & 1.7 $\pm$ 0.3 & 24.1 $\pm$ 0.8 \\
      PDR 3 & 23 $\pm$ 3 & 1.7 $\pm$ 0.4 & 24.6 $\pm$ 0.8 \\
      PDR 4 & 22 $\pm$ 3 & 1.5 $\pm$ 0.3 & 22.3 $\pm$ 0.7 \\
      PDR 5 & 22 $\pm$ 3 & 1.5 $\pm$ 0.3 & 22.9 $\pm$ 0.7 \\
      PDR 6 & 24 $\pm$ 3 & 1.8 $\pm$ 0.3 & 26.0 $\pm$ 0.9 \\
      PDR 7 & 25 $\pm$ 3 & 1.7 $\pm$ 0.3 & 26.9 $\pm$ 1.0 \\
      PDR 8 & 26 $\pm$ 4 & 1.7 $\pm$ 0.4 & 28.6 $\pm$ 1.2 \\
      UCHII & 29 $\pm$ 5 & 1.8 $\pm$ 0.3 & 32.7 $\pm$ 1.5 \\
      \hline\hline
    \end{tabular}
  \end{minipage}
\end{table}

To determine the dust temperature structure of \mia, we performed aperture
photometry measurements on selected areas in the field.
The apertures are shown in Fig. \ref{fig-sh104-main} and sample the interior
of the bubble and the photodissociation region (PDR), including the UC\hii\
region associated with IRAS~20160+3636. We used a single aperture (not shown) to
account for the background emission.
We fitted the PACS+SPIRE emission for all regions. These data represent the cold
emission component, therefore we fitted a single temperature. We did this first
allowing $\beta$ to vary, and later fixing $\beta=1.5$. The resulting
$T_{dust}$ and $\beta$ values are shown in Table \ref{table-beta}, their
uncertainties are the formal $1\sigma$ values from the fitting procedure.
Figure \ref{fig-seds} shows examples of the fitted spectral energy distributions
(SEDs) for the regions \textit{UCHII} and \textit{PDR 1}. Apart from the clear
difference in $T_{dust}$, it can be seen that for \textit{UCHII} the PACS+SPIRE
emission does not allow to sample the peak of the SED, which is reflected in a
larger error in the fit.
This is also seen in the four \textit{Interior \#} regions, suggesting the
presence of a warmer emission component that contributes to the
shorter-$\lambda$ emission. Observations in the mid-IR can provide a constraint
on this component, and a 2-temperatures fitting would be more appropriate to
determine the $T_{dust}$ and $\beta$ values.

The temperatures throughout the PDR are between $\sim20$ and $\sim30$\,K, and
the UC\hii\ region is marginally warmer. The region outside the
bubble (region \textit{Outer}) is the coldest, while the regions mapping the
interior of the bubble are hotter, with an average temperature of $\sim40$\,K.
The average $\beta$ obtained is $\sim1.5$, therefore the temperatures obtained
from the fit with a fixed $\beta=1.5$ do not significantly differ from those
obtained with a free $\beta$ (see Table \ref{table-beta}).

\begin{figure}[t]
  \centering
  \includegraphics[width=\columnwidth]{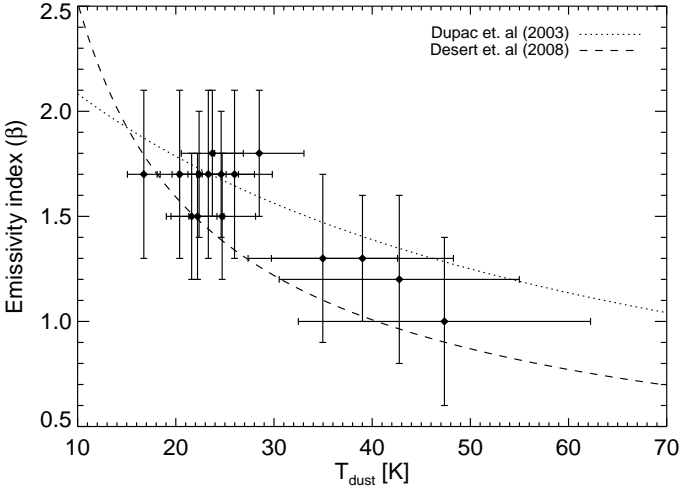}
  \caption{Distribution of the emissivity index $\beta$ vs. dust temperature.
The dotted and dashed lines are the relationships found by \citet{dupac2003}
and \citet{desert2008}, respectively. We cannot distinguish
between the two relationships, but lower $\beta$ values correspond to
 higher temperatures (see text).}
  \label{fig-beta}
\end{figure}

Figure \ref{fig-beta} shows the distribution of the spectral indices $\beta$
vs. dust temperature, along with the relationships found by \citet{dupac2003}
and \citet{desert2008}. Within the uncertainties, our results agree with both
relationships. We can also identify two groups, one with temperatures
between $35$ and $47\,$K and an emissivity index between $1.0$ and $1.3$, and
the other with $T = 17 - 29\,$K and $\beta=1.5\, - 1.8$. These two groups show
that on average higher $\beta$ values are preferentially associated with colder
material.

Although anti-correlation between $\beta$ and $T_{dust}$ is
reported in the literature (e.g., \citealt{dupac2003,yang2007,desert2008}), it
is yet not clear which physical processes are behind it.
Other authors studying this relationship examine the emission from
different regions scattered in the sky. In contrast, we are finding this
anti-correlation in the analysis of one contiguous complex object at
a specific location in the sky. A $\beta-T_{dust}$ relationship may
indicate a change of the dust properties \citep{stepnik2003}. Grain fluffiness
in particular increases the emissivity index while keeping a relatively low
temperature (e.g., \citealt{stognienko1995,fogel1998}).

Fluffy grains result from grain coagulation and growth.
The grain coagulation timescale and feasibility depend on factors
such as the existence of ice mantles, grain size, and relative grain
velocities.

We are finding the highest $\beta$ values and lowest $T_{dust}$
values toward the PDR of \mia, which would imply then that the fluffiest and
largest grains are located in the PDR. The question remains whether the PDR
dust coagulated after the creation of the \hii\ region, or if the birth of the
\hii\ region has destroyed the already coagulated dust located in the ionized
cavity.

\section{Gas properties}

\begin{figure}[t]
  \centering
  \includegraphics[width=\columnwidth]{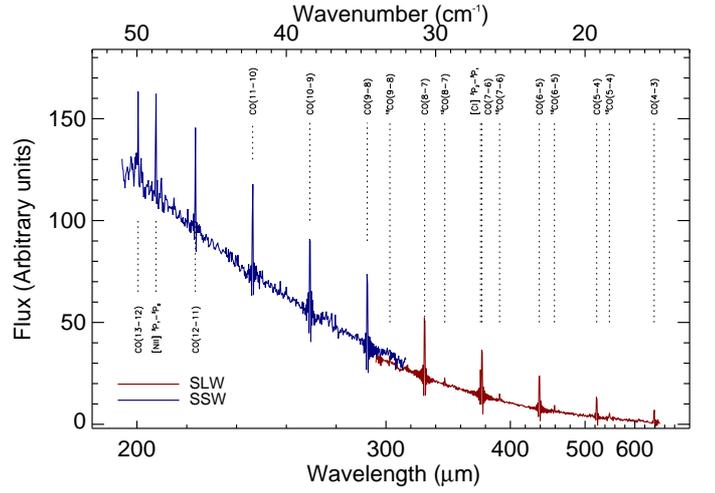}
  \caption{SPIRE-FTS spectrum corresponding to the central pixels of SLW
(pixel C3, red) and SSW (pixel D4, blue), towards the UC\hii\ region associated
with IRAS~20160+3636 (pointing \textit{E}). The lines detected are labeled. The
jump between continuum levels is a calibration effect.}
  \label{fig-spectrum}
\end{figure}

The seven pointings of the central pixels observed with SPIRE-FTS are marked
with
red (SLW) and blue (SSW) circles in Fig.~\ref{fig-sh104-main}. In total we
detected the transitions described in Table~\ref{table-spectra}.
The richest spectrum (Fig.~\ref{fig-spectrum}) was obtained towards pointing
\textit{E}, which targets the UC\hii\ region associated with IRAS~20160+3636.
The most prominent features are the CO \mbox{$J$-ladder} and the \nii\
transition, an
important ionized-gas coolant and a proxy for the H$\alpha$ flux (see e.g.,
\citealt{oberst2006}).

\begin{figure}[h]
  \centering
  \includegraphics[width=\columnwidth]{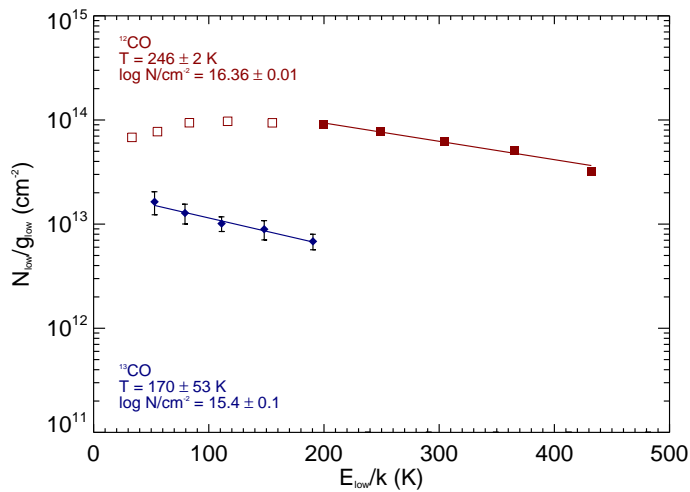}
  \caption{$^{12}$CO (red squares) and $^{13}$CO (blue diamonds)
rotational diagram for pointing \textit{E} towards the UC\hii\ region. The
solid lines are the best linear fit. Points included in the fit are shown as
filled, points excluded as open. The turnover into the optically thick regime is
noted in the lower-level transitions of $^{12}$CO.
}
  \label{fig-rot}
\end{figure}

In the simple hypothesis of optically thin emission, we plotted $^{12}$CO
and $^{13}$CO excitation diagrams, following the formulation of
\citet{johansson1984}. An example is shown in Fig. \ref{fig-rot} for pointing
\textit{E}. The slope of the linear fit for $^{12}$CO results in a temperature
of $T^{12} = 246\pm2\,$K, and $T^{13} = 170\pm53\,$K for $^{13}$CO. The total
column densities derived are $N_T^{12} = 10^{16.36\pm0.01}$ and $N_T^{13} =
10^{15.4\pm0.1}\,{\rm cm}^{-2}$, respectively.
The fits in Fig. \ref{fig-rot} do not include all the measurements. The
downturn seen in the lower levels of $^{12}$CO is interpreted
as the optically thick/thin regime turnover, and is most likely a real physical
feature and not an instrumental or calibration effect, because it is only seen
for
that species and not, for example, for $^{13}$CO. If the lines were emitted from
an optically thick medium, their intensity would be underestimated, thus their
respective column densities would also be a lower limit.
Therefore, the optically thin assumption would hold for $^{12}$CO only for
transitions higher than $J=9\rightarrow8$, and those are the points included in
the fit.

For $^{13}$CO on the other hand only the lines with
wavelengths in the SPIRE-FTS SLW range are detected. These correspond to the $J
= 5\rightarrow4$ to $J = 9\rightarrow8$ transitions. The $^{13}$CO lines in
the SPIRE-FTS SSW wavelenght range ($J = 10\rightarrow9$ to $J =
14\rightarrow13$) are detected as upper limits, because the line positions are
found displaced from their expected positions, indicating that we are probably
seeing some ``outlier'' noise features rather than the lines themselves.
Therefore, we fit and show in Fig. \ref{fig-rot} only the five
$^{13}$CO transitions detected with SPIRE-FTS SLW.
Following the reasoning of the previous paragraph, the $^{13}$CO transitions are
most likely optically thin.

\begin{table}[b]
  \renewcommand{\arraystretch}{1.1}  
  \begin{minipage}{\columnwidth}
    \centering
    \caption{Lines detected with SPIRE-FTS in the different pointings.}
    \scriptsize
    \label{table-spectra}
    \renewcommand{\footnoterule}{}
    \begin{tabular}{l|cc||l|cc}
      \hline\hline
      Transition & Rest $\lambda$ ($\mu$m) & E$_{up}$ (K) & Transition & Rest
$\lambda$ ($\mu$m) & E$_{up}$ (K) \\
      \hline
      CO$(4-3)$ & $650.3$ & $55.4$ & $^{13}$CO$(5-4)$ & $544.2$ & $79.4$ \\
      CO$(5-4)$ & $520.2$ & $83.0$ & $^{13}$CO$(6-5)$ & $453.5$ & $111.1$ \\
      CO$(6-5)$ & $433.6$ & $116.3$ & $^{13}$CO$(7-6)$ & $388.7$ & $148.2$ \\
      CO$(7-6)$ & $371.7$ & $155.0$ & $^{13}$CO$(8-7)$ & $340.2$ & $190.5$ \\
      CO$(8-7)$ & $325.2$ & $199.3$ & $^{13}$CO$(9-8)$ & $302.4$ & $238.1$ \\
      CO$(9-8)$ & $289.1$ & $249.1$ & \ci & $609.1$ & $23.6$ \\
      CO$(10-9)$ & $260.2$ & $304.4$ & \cia & $370.4$ & $47.3$ \\
      CO$(11-10)$ & $236.6$ & $365.3$ & CH$^+(1-0)$ & $359.0$ & $40.1$ \\
      CO$(12-11)$ & $216.9$ & $431.7$ & \nii & $205.2$ & $70.2$ \\
      CO$(13-12)$ & $200.3$ & $503.6$ & & & \\
      \hline\hline
    \end{tabular}
  \end{minipage}
\end{table}

The two distinct gas temperatures obtained with $^{12}$CO and $^{13}$CO
suggest two different gas components or a stratification of
the emitting region. The colder component is traced by the optically-thin
$^{13}$CO transitions at energy levels for which $^{12}$CO is optically thick,
while the hotter component is traced by the more energetic, optically-thin
$^{12}$CO transitions. Therefore it is likely that the colder gas is located at
grater depths in the PDR than the hotter gas.

Assuming similar emitting volumes and beam filling factors as well
as optically thin emission, our CO and
$^{13}$CO column density values imply an abundance ratio ${\rm
[^{12}CO]/[^{13}CO]}$, which is several times lower than the reported elemental
value of ${\rm ^{12}C/^{13}C}\sim69$ \citep{wilson1999}. This would imply
an enhancement of the $^{13}$CO isotopologue abundance.

However, several factors can contribute to the low ${\rm ^{12}C/^{13}C}$
abundance ratio we find. Perhaps the most important one would be the assumption
of optically thin emission for $^{12}$CO. We used the high-energy
transitions to derive its column density, and although in a first analysis they
appear to be optically thin, it might not be the case (see e.g.,
\citealt{habart2010}). To address this issue we will
present and analyse PDR models of the \mia\ region in a forthcoming paper.

\section{Summary}

With \textit{Herschel} PACS and SPIRE data we have analysed dust and gas
properties of the bubble-shaped \hii\ region \mia.
Aperture photometry of PACS+SPIRE images allowed us to derive the dust
emissivity index $\beta$ and the dust temperature throughout \mia. We found 
two different groups, one at colder temperatures and higher $\beta$, and the
other at warmer temperatures and lower $\beta$. We recover the inverse $\beta
- T_{dust}$ relationship reported in the literature, but the uncertainty
in our fit prevents us from finding a precise depence. As an example, the
different relationships found by \citet{dupac2003} and \citet{desert2008} are
both compatible with our results.
PACS and SPIRE have allowed us to constrain the ``cold'' side of the SED. We
have estimated both $\beta$ and $T_{dust}$ simultaneously, but the
uncertainties remain relatively high. With the data at hand we can only
fit a single temperature component, which disregards the contribution of warmer
components to the PACS shorter wavelengths.
The existance of a $\beta - T_{dust}$ anti-correlation could be due to
differences of the dust grain properties between the PDR and the \hii\ cavity.

SPIRE-FTS spectra at different pointings throughout \mia\ have unveiled the CO
chemistry of the region in more detail. We detect the $^{12}$CO and $^{13}$CO
$J$-ladders up to the $J = 13\rightarrow12$ and $J = 9\rightarrow8$ transitions
respectively, revealing the warm gas component in the region.
Rotational diagrams towards the UC\hii\ region in the PDR of \mia\ show that the
$^{13}$CO emission is optically thin and also the $^{12}$CO for transitions
above the $J=8$ level. The emission shows
two different gas components, a colder one with a temperature of $\sim170\,$K
and a hotter one at a temperature of $\sim250\,$K.
The CO column densities derived would suggest an enhancement of the
$^{13}$CO
isotopologue abundance ratio with respect to the elemental value, but the
uncertainties of the different assumptions are still too large to confirm that
result.
In a follow-up paper we will show models of the PDR of \mia, which will provide
better constraints on the gas temperature, density and column density
structure.

\begin{acknowledgements}

SPIRE has been developed by a consortium of institutes led by Cardiff Univ. (UK)
and including Univ. Lethbridge (Canada); NAOC (China); CEA, LAM (France); IFSI,
Univ. Padua (Italy); IAC (Spain); Stockholm Observatory (Sweden); Imperial
College London, RAL, UCL-MSSL, UKATC, Univ. Sussex (UK); Caltech, JPL, NHSC,
Univ. Colorado (USA). This development has been supported by national funding
agencies: CSA (Canada); NAOC (China); CEA, CNES, CNRS (France); ASI (Italy);
MCINN (Spain); Stockholm Observatory (Sweden); STFC (UK); and NASA (USA).
PACS has been developed by a consortium of institutes led by MPE 
(Germany) and including UVIE (Austria); KUL, CSL, IMEC (Belgium); CEA, 
LAM (France); MPIA (Germany); IFSI, OAP/AOT, OAA/CAISMI, LENS, SISSA
(Italy); IAC (Spain). This development has been supported by the funding 
agencies BMVIT (Austria), ESA-PRODEX (Belgium), CEA/CNES (France),
DLR (Germany), ASI (Italy), and CICT/MCT (Spain).
Part of this work was supported by the ANR (\emph{Agence Nationale pour la
Recherche}) project ``PROBeS'', number ANR-08-BLAN-0241.

\end{acknowledgements}


\end{document}